\documentclass[conference]{IEEEtran}
\IEEEoverridecommandlockouts

\usepackage{cite}
\usepackage{amsmath,amssymb,amsfonts}
\usepackage{algorithmic}
\usepackage{graphicx}
\usepackage{textcomp}
\usepackage{xcolor}

\usepackage{url}
\usepackage{xcolor}

\def\BibTeX{{\rm B\kern-.05em{\sc i\kern-.025em b}\kern-.08em
    T\kern-.1667em\lower.7ex\hbox{E}\kern-.125emX}}
\begin{document}

\title{Intelligent Software System for Low-Cost, Brightfield Segmentation: Algorithmic Implementation for Cytometric Auto-Analysis\\

\thanks{The research was supported by ITMO University Research
Projects in AI Initiative (RPAII) (project \#640103, Development of methods for automated processing and analysis of optical and atomic force microscopy images using machine learning techniques).}
}

\author{\IEEEauthorblockN{Surajit Das}
\IEEEauthorblockA{\textit{Infochemisry Scientific Center} \\
\textit{ITMO Unniversity}\\
Saint Petersburg, Russia \\
mr.surajitdas@gmail.com}
\and
\IEEEauthorblockN{Pavel Zun}
\IEEEauthorblockA{\textit{Infochemisry Scientific Center} \\
\textit{ITMO Unniversity}\\
Saint Petersburg, Russia \\
pavel.zun@gmail.com}
}

\maketitle

\begin{abstract}
Bright-field microscopy, a cost-effective tool for live-cell culture, is often the only imaging resource available—along with standard CPUs—for low-budget labs. The inherent challenges of bright-field images—noise, low contrast, and dynamic morphology—combined with limited GPU resources and complex software interfaces, hinder research output. This article presents a novel microscopy image analysis framework for CPU-based desktops in low-resource labs. The Python program enables cytometric analysis of live, unstained cells through an advanced computer vision and machine learning pipeline. Operating entirely on label-free data, it requires no manual annotations or training. The framework features a user-friendly, cross-platform GUI needing no programming skills, plus a scripting interface for programmatic control and integration. Its end-to-end workflow performs semantic and instance segmentation, feature extraction, analysis, evaluation, and automated report generation. The modular design ensures easy maintenance, flexible integration, and both single-image and batch processing. Validated on multiple unstained cell types from the LIVECell dataset, the framework shows higher accuracy and reproducibility than Cellpose and StarDist. Its competitive CPU segmentation speed underscores its potential for basic research and clinical applications—especially in cell transplantation for personalized medicine and muscle regeneration. Access to the application is provided for reproducibility \footnote{This paper is under peer review. We shall release the code soon. 
}.
\end{abstract}


\section{Introduction}
Microscopy image analysis has undergone a paradigm shift with the introduction of artificial intelligence (AI) and machine learning (ML). However, most AI-based solutions rely on large annotated datasets, extensive GPU resources, and strong contrast or staining, which limit their usability in live-cell and label-free microscopy. Additionally, most of the tools facilate script-based operation or complex interface causing hurdles to non-coder. This motivates the development of lightweight, explainable, and training-free methods for biological laboratories lacking computational infrastructure.

This paper presents an end-to-end, semi-unsupervised segmentation framework based on the recently introduced Homogeneous Image Plane (HIP) model~\cite{das2024hip}. The proposed system operationalizes the algorithm through an intuitive graphical user interface (GUI), bridging theory and practical deployment. The work directly extends \cite{das2024hip} by embedding its algorithmic components into a reproducible, modular tool for biomedical researchers. Hence, the article outlines the core features\footnote{The algorithmic discussion, accompanied by detailed comparative results with the SOTA model, is incorporated in the referred article by the authors.} of an integrated Python-based image analysis GUI application for biomedical research or clinical practice, focusing on cellular and morphological characterization. It provides image inspection, processing, and quantitative analysis through a graphical user interface (GUI).

\textbf{Image Management and Visualization: }
The tool supports standard formats (PNG, JPG, BMP, TIFF) with an interactive canvas for scrolling, zooming, and background opacity adjustment for overlays. Users can define regions-of-interest (ROI) for independent analysis.

\textbf{Interactive Pixel and Spatial Analysis: }
Real-time cursor tracking displays XY coordinates. Pixel inspection retrieves RGB values and grayscale conversions. User-defined neighborhood kernels enable first-order (mean, median, mode, SD), higher-order (skewness, kurtosis), and spatial statistics (Moran’s I, variogram measures). Results appear dynamically in a scrollable widget.

\textbf{Modular Image Processing Pipeline: }
The GUI links to a modular workflow focused on segmentation:
\texttt{(I) Semantic Segmentation:} Primary segmentation (`HIP Actual Grayscale Analysis') with configurable parameters (shift, span, masking threshold, contrast threshold, randomness index).
\texttt{(II) Post- Segmentation:} Refinement via denoising/cleaning with morphological operations.
\texttt{(III) Instance Segmentation:} Algorithms for separating individual objects (e.g., cells).
\texttt{(IV) Auto Analysis \& Reports:} Extraction of features and auto-generation of reports.

\textbf{Comprehensive Quantitative Analysis: }
Users can compute a wide range of metrics: \textbf{General} (object count, cell density); \textbf{Geometry} (perimeter, area, aspect ratio); \textbf{Morphology} (roundness, shape, convexity); \textbf{Localization} (2D coordinates, centroid); \textbf{Orientation} (object orientation); and \textbf{Voronoi Entropy} (spatial disorder from Voronoi tessellation).

\textbf{Data Handling and Reporting: }
Pixel statistics and analysis results export to Excel. The `Report' function generates analytical profiles with a progress bar for long tasks.

\textbf{Profile-Based Batch Processing: }
Profiles store reusable parameters for single or batch processing, ensuring reproducibility. Batch mode produces analytical reports with statistical summaries and visualizations, suitable for research and clinical use.

Our novelty lies in the following contributions: The algorithmic implementation of the semantic segmentation and post-segmentation de-noising methods originally proposed by Das \& Zun, integrated effectively into a system software framework. The development of novel modules for instance segmentation and automatic report generation, forming a comprehensive end-to-end pipeline for cytometry analysis. The design and creation of a user-friendly GUI-based framework that enhances accessibility and usability for end users.

The rest of the paper is organised with the sections Literature Review, Methodology Overview, System Implementation, Results, Discussion and Conclusion.

\section{Literature Review}
Over the past two decades, microscopy image analysis has progressed from manual thresholding to AI-driven methods, enhancing accuracy, reproducibility, and throughput. Early tools like CellProfiler (2006) \cite{ref_cellprofiler}, ImageJ/Fiji \cite{ref_fiji}, and Icy (2011) \cite{ref_icy} combined classical image processing with basic ML for segmentation and tracking but relied on staining and high-contrast images. Also, the tools rely
heavily on manual parameter tuning.

The 2010s saw ML-based tools such as Ilastik (2011) \cite{ref_ilastik}, CellCognition (2010) \cite{ref_cellcognition}, and TrackMate (2017) \cite{ref_trackmate}, enabling pixel classification, supervised learning, and improved tracking. Feature engineering was labor-intensive, and performance varied with image quality.

Deep learning (2018–present) transformed segmentation. DeepCell (2016) \cite{ref_deepcell}, Cellpose (2020) \cite{ref_cellpose}, and StarDist (2018) \cite{ref_stardist} achieved near-human accuracy. Cellpose, using a modified U-Net, predicts horizontal and vertical flow fields along with binary cell masks, which are post-processed for precise instance segmentation, even for irregularly shaped cells. ZeroCostDL4Mic (2020) \cite{ref_zerocostdl4mic} enabled DL without coding, though requiring large datasets and high computation.

Recent tools like Omnipose (2022) \cite{ref_omnipose} and MISIC (2023) \cite{ref_misic} handle complex shapes, unstained cells, and sparse label conditions. Omnipose excels in bacterial and irregular cells; MISIC in stain-free and sparse-label scenarios. Emerging trends include hybrid classical-DL models, Semi-supervised learning, and user-friendly GUIs. Remaining gaps include domain-specific analysis for live cells (e.g., myoblasts), multi-modal integration, computational cost, accessibility and interface complexities.

{\color{black}Additionally, many models behave as black boxes and require large annotated datasets, limiting transparency and generalisability. In contrast, the proposed fuzzy-statistical approach retains interpretability through explicit rule-based intensity reasoning and operates effectively without labeled data.}

Recently, S. Das \& P. Zun developed a semi-unsupervised algorithm for semantic segmentation based on the notion of a homogeneous image plane \cite{das2024hip} which fulfils most of the gaps mentioned above. A significant gap in their research is the lack of a complete, end-to-end implementation for cytometric analysis. Our work complements theirs by adding modules for instance segmentation and an automated report-analysis system. Another important gap is that such algorithms are typically not integrated into an end-to-end AI/ML pipeline but are reported as isolated entities. This demands the involvement of skilled computer practitioners to use them.

Therefore, the proposed framework overall addresses the abovementioned gaps via fuzzy logic and spatial statistics in a Python-based, user-friendly tool for end-to-end cytometric auto-analysis in regenerative medicine. The framework has been successfully tested on a span of cell types.

\section{Methodology}
The core algorithm is adapted from Das et al. work~\cite{das2024hip}, which introduced the concept of a Homogeneous Image Plane and the Spatial Standard Deviation from Local Mean (SSDLM) metric for background separation. The key components are:

\begin{itemize}
    \item \textbf{Spatial Standard Deviation from Local Mean (SSDLM):} Estimates homogeneity by measuring local intensity variation.
    \item \textbf{Cumulative Squared Shift of Nodal Intensity (CSSNI):} Quantifies texture-driven local fluctuations.
    \item \textbf{Fuzzy Logic System:} Handles uncertainty in pixel classification via three membership functions (dark, gray, bright).
    \item \textbf{Adjusted Variogram:} Captures global spatial variation normalized by Euclidean distance.
\end{itemize}
These components together form a rule-based adaptive segmentation pipeline without the need for training or labels. Furthermore, the authors have designed and implemented two additional modules focused on instance segmentation and the automatic generation of reports, utilising univariate and bivariate analyses of features extracted automatically.\\

\begin{figure}[htbp]
\centerline{\includegraphics[width=0.48\textwidth]{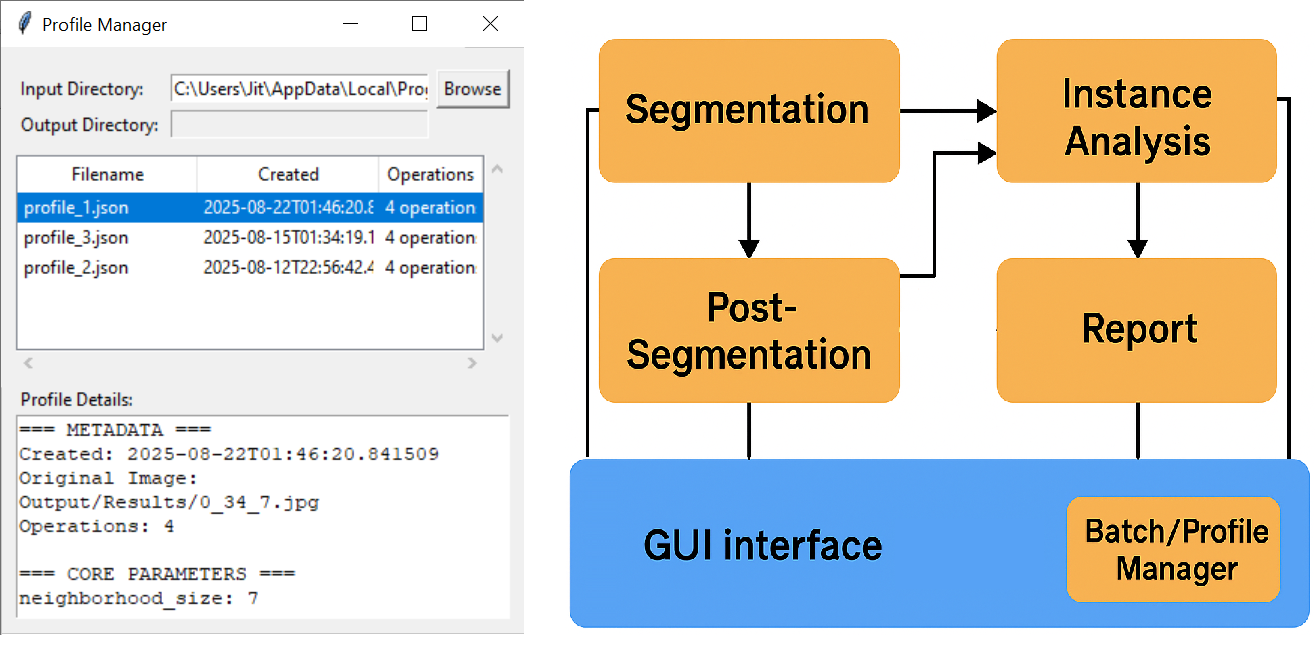}}
\caption{(Left) Profile Manager for batch image segmentation, displaying stored parameter configurations for reproducible analysis across sessions or datasets. (Right) Schematic diagram showing the connected modules of the microscopy segmentation application}
\label{moduler_coonect}
\end{figure}

The GUI is built in Python using Tkinter and OpenCV, optimized with Numba for CPU efficiency. It consists of four modules: segmentation, post-processing, instance analysis, and automated reporting. Each module can be executed individually through the GUI or scripting interface. The moduler connectivities are represented in Fig ~\ref{moduler_coonect}.

\subsection{Segmentation}

\begin{figure}[htbp!]
  \centering

    \centering
    \includegraphics[width=0.48\textwidth]{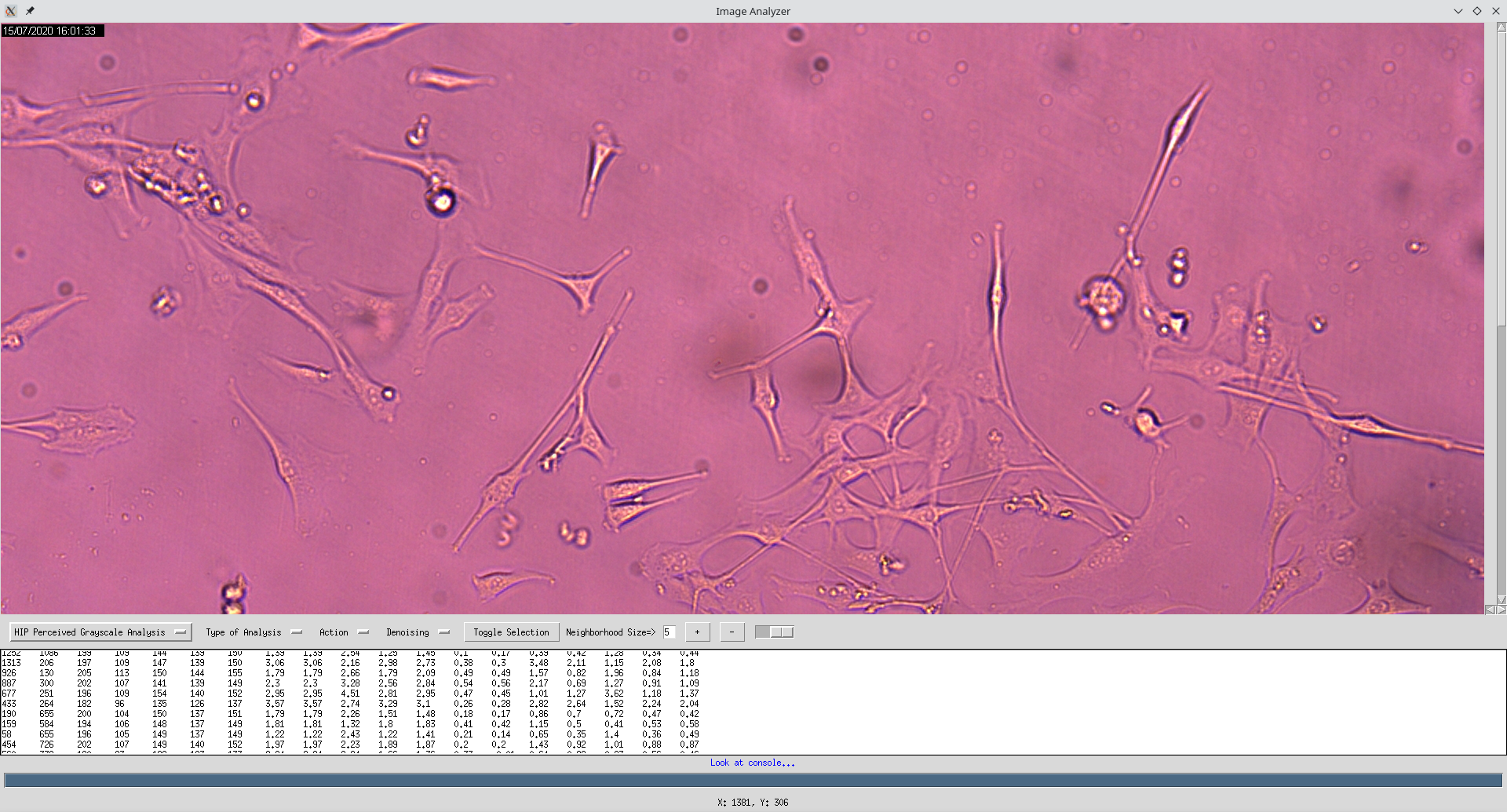}
  \caption{Graphical interface for selecting background pixels within a configurable neighborhood (e.g., 5×5). The window allows users to sample pixel intensities interactively, enabling adaptive thresholding and homogeneity-based segmentation calibration $5 \times 5$}

     \label{cal_gui_seg}
\end{figure}

\begin{figure}[htbp!]
  \centering

    \centering
    \includegraphics[width=0.48\textwidth]{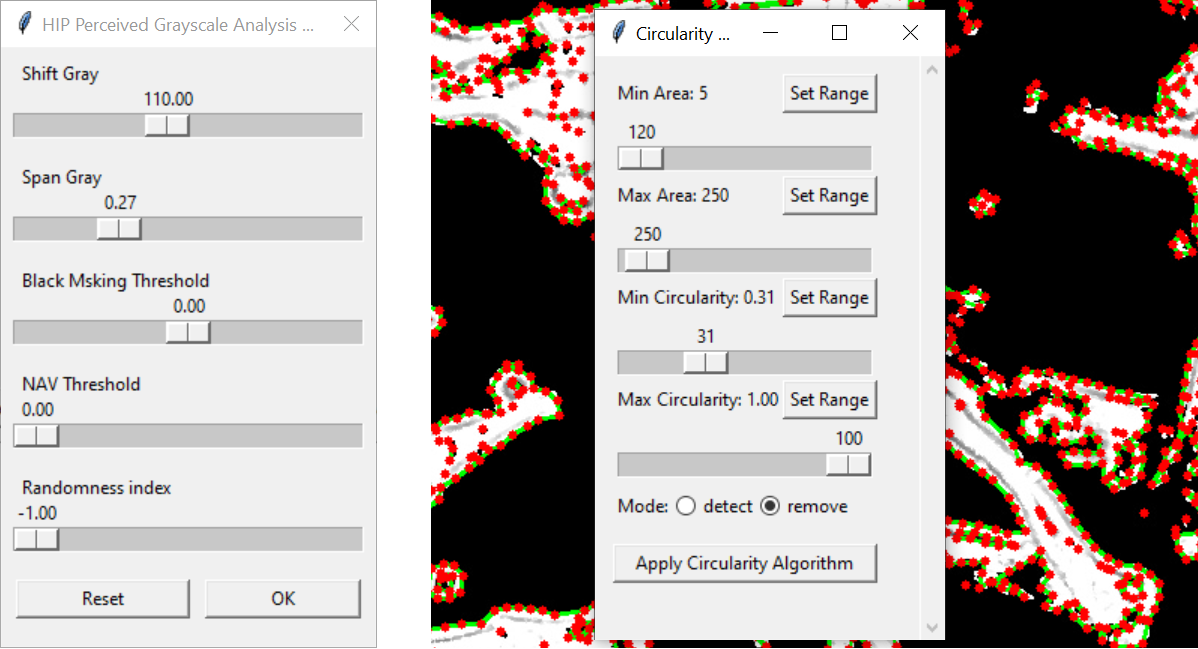}
    \caption{(Left) Calibration module for adjusting brightness, masking, and randomization thresholds prior to segmentation.
(Right) Denoising module applying morphological filters (erosion, dilation, median blur) to enhance contour definition and reduce noise artifacts in unstained cell images.}

  \label{fig:cal_denois}
\end{figure}

The semantic segmentation module applies neighborhood-based statistical filtering and fuzzy intensity transformation to separate texture (cells) from background. It integrates Numba-optimized functions for efficiency, using local statistics (mean, median, mode, skewness, kurtosis), spatial measures (Moran’s I, variogram, adjacency shift), and fuzzy logic for pixel classification. Additional steps include homogeneous masking, circle removal via Hough transform, and partially adaptive thresholds (sv1–sv5) etc. Processing is performed with multi-channel support, producing binary masks that serve as inputs for post-processing and instance analysis. {\color{black}Users can execute the segmentation task either through clicking GUI (Fig~\ref{cal_gui_seg})  or via the console commands. Users are able to do manual calibration if required before starting segmentation by adjusting the sliders "Shift Gray", "Span Gray", "Black Masking Threshold", "NAV Threshold", "Randomness Index", etc. which are illustrated in the calibration window (left part of Fig~\ref{fig:cal_denois}).}

\subsection{Post-Segmentation Processing Module}

The module provides a comprehensive suite for refining and denoising segmented images.  

\textbf{Core Architecture:} Implemented as a \texttt{ProfileProcessor} class managing the post-processing pipeline. Key data structures include: image data (BGR), contour list, operations history, and current profile. SQLite ensures persistent storage of operation history with timestamps, parameters, and performance metrics for auditability.  

\textbf{Input/Output Handling:} Supports loading pickle files with images, contours, histories, and profiles. Output is saved as both pickle (for further processing) and PNG (for visualization), with proper color space conversion between grayscale and BGR.  

\textbf{Contour Processing and Analysis:} Enables contour detection and management with intelligent sampling (configurable spacing, Euclidean-distance-based point reduction, structured sampling) while preserving morphological integrity. Fully compatible with OpenCV contour functions.  

\textbf{Morphological Operations:}  
- \textit{Area-based Filtering:} Eliminates regions below minimum area, with modes for removal (filling) or marking (visualization); supports hierarchical contour processing.  
- \textit{Shape-based Filtering:} Aspect ratio elimination (customizable bounds), circularity filtering (isoperimetric quotient), eccentricity-based filtering for elliptical structures.  
- \textit{Image Enhancement:} Median blur (auto kernel size), erosion/dilation (configurable kernels), and threshold-based filling for artifact removal.  

\textbf{Advanced Contour Refinement:} Includes Sobel gradient edge detection, distance transforms, force-based contour deformation (parameters: force constant, decay exponent, gradient threshold), convex hull enforcement, and multi-stage smoothing via neighborhood averaging.  

\textbf{Parameter Normalization System:} Provides explicit mapping of parameter names, automatic key normalization, numeric type conversion, and flexible matching strategies for consistent operation handling.  

\textbf{Operation Mapping and Execution:} Supports multiple naming conventions, case-insensitive and substring-based matching, parameter validation, adaptive execution, and robust error handling with fallbacks.  

\textbf{Quality Control and Metrics:} Features exception handling with detailed reporting, success/failure tracking, performance metrics recording, and full processing history maintenance.  
This module offers a complete solution for post-segmentation refinement and analysis of biomedical images, particularly for cellular morphology studies requiring precise contour characterization and filtering for accurate quantitative analysis. Users can execute "Post-Segmentation Processing" either through the GUI (via a click) or via the console commands.

\subsection{Instance Segmentation \& Morphological Analysis Module}

This module automates detection, segmentation, and quantitative analysis of individual objects, focusing on cellular structures.

\textbf{Core Functionality:} Performs instance segmentation via contour-based detection, extracting features and generating annotated visuals and quantitative data.

\textbf{Input Processing:} Accepts images in multiple formats via pickle serialization (grayscale, single-channel BGR, three-channel BGR). Automatic color conversion ensures consistency; adaptive thresholding generates binary masks for contour detection.

\textbf{Contour Detection and Object Identification:} Uses OpenCV contour detection with , configurable minimum area thresholds remove artifacts, and each object receives a unique identifier.

\textbf{Morphological Feature Extraction:} Autonomously computes geometric features (area, perimeter, aspect ratio, bounding box), morphological metrics (roundness, shape classification, convexity, solidity), spatial/orientation features (centroid, top-left coordinates, orientation angle, eccentricity), and advanced metrics (Voronoi entropy, cell density, equivalent diameter, extent).

\textbf{Visual Output Generation:} Produces annotated, bounding-box, and contour images with configurable annotation parameters (font scale, color, thickness).

\textbf{Data Export and Reporting:} Exports structured Excel spreadsheets with all features, automatic naming, console summaries, real-time feedback, and debugging support.

\textbf{Configuration and Customization:} Users can adjust feature selection, minimum area thresholds, annotation appearance, and output management (file naming, directories).

\textbf{Scientific Image Analysis:} Enhances extended morphological characterization supporting intensity-based features, advanced convex hull metrics, refined eccentricity and solidity, and principal axis analysis for orientation.

\textbf{Performance and Reliability:} Robust error handling, input validation, adaptive processing, memory-efficient contour handling, and cross-platform compatibility.
This module provides a reliable solution for automated cellular analysis, combining quantitative data with visual verification for biomedical morphological studies.

\subsection{Automated Report Generation Module}

This module produces comprehensive HTML reports from Excel-based cellular analysis data discussed above section (Morphological Feature Extraction), completing the image analysis pipeline by converting quantitative morphological measurements into structured, presentable outputs.

\textbf{Core Functionality:} Automatically generates HTML reports with statistical summaries, distribution visualizations, structured analytical results, and professional formatting for research documentation.

\textbf{Architecture:} Built on \texttt{jitprofiler} with modular separation of data processing and presentation layers.

 \textit{jitprofiler} is a Python library developed by the author and publicly available on the Python Package Index (PyPI). \textit{jitprofiler} is designed to transform raw datasets into comprehensive, interactive analytical reports. The library employs a multi-layered analytical methodology that begins with robust data quality assessment and statistical characterization, then generates interactive visualizations using the Plotly framework. These dynamic charts—including histograms, box plots, scatter plots, and correlation heatmaps—enable researchers to engage directly with the data through zooming, hovering, and panning functionalities, transforming static analysis into an interactive discovery process, enabling researchers to engage directly with the data’s underlying structure and anomalies. 

The statistical characterization performed by \textit{jitprofiler} encompasses descriptive, summary, and quartile statistics to quantify central tendency and variability (mean, standard deviation, minimum, maximum, percentiles, skewness, and kurtosis). It further integrates distributional analysis through histograms and boxplots to visualize spread and outliers, and correlation analysis using Pearson’s coefficient and heatmaps to reveal inter-feature relationships such as the strong Perimeter–Area association. Additionally, variability and dispersion measures (IQR, standard deviation), extreme value detection for rare morphological instances, and frequency analysis to confirm data uniqueness are included. Together, these analyses provide a complete quantitative understanding of morphological heterogeneity and feature coherence within the cellular dataset. It provides several other functions and handles categorical values automatically. Overall, the automated cytometric report-generating module accurately quantifies morphometric variability within the sample, producing statistically consistent and biologically interpretable metrics.

The system synthesizes these interactive visualizations with comprehensive statistical profiling into a unified HTML dashboard featuring toggle-based section control for focused analysis. The entire report maintains academic rigor while offering export capabilities to PDF for publication purposes. As an open-source tool, \texttt{jitprofiler} represents an advancement in accessible data science workflows, providing researchers across domains with standardized, reproducible analytical procedures that accelerate exploratory data analysis while ensuring methodological consistency and deepening intuitive understanding of complex datasets.

\textbf{Inputs:} Excel files containing geometric, morphological, and spatial measurements, along with configuration parameters (image identifiers and processing settings).

\textbf{Outputs:} Professional HTML documents with structured layouts, integrated charts, descriptive statistics, and cross-platform compatibility.

\textbf{File Management:} Standardized naming conventions and directories embed image identifiers, neighborhood size, timestamps, or version indicators for traceability.

\textbf{Pipeline Integration:} Consumes processed data from segmentation and feature extraction, maintaining parameter consistency and ensuring quality through standardized reporting.

\textbf{Error Handling:} Validates file existence and format, checks parameter integrity, and provides informative messages.
where \verb|\<input\_image\_path>| is the source image and \verb|\<neighborhood\_size>| feature extraction parameter.

\textbf{Quality Assurance:} Ensures input completeness, parameter-result consistency, standardized formatting, and cross-referencing for multi-image studies.

The module transforms raw cellular data into actionable insights and polished research documentation. Users can execute Report Generation task either through the GUI (via a click) or via the console commands.

\subsection{Profile Management and Batch Processing System}

The profile manager (Fig~\ref{moduler_coonect}) allows researchers to save, reuse, and share complete image processing configurations, ensuring reproducibility and consistency across experiments. Profiles are stored in a standardized JSON format with metadata (creation time, source image, version) and all analysis parameters. A graphical interface provides an interactive browser for managing profiles, previewing settings, and configuring directories, while the batch engine automates large-scale processing with real-time progress updates, error handling, and cancellation options.  

From the user’s perspective, workflows are straightforward: create a profile once, apply it to single images or entire datasets, and receive consistent results with automated reporting. The system manages input/output directories automatically, organizes results into structured folders, and preserves original naming for easy traceability.  

Key benefits include reproducibility (version-controlled profiles and audit trails), scalability (batch execution on large collections or time-series data), and flexibility (parameter optimization, protocol standardization, or quality control). Built-in error recovery, efficient memory handling, and detailed logging ensure reliability even in high-throughput studies.  

In practice, the system streamlines experimental pipelines: users define their analysis settings once, let the system run across thousands of images, and obtain consistent, validated outputs with minimal manual intervention.

\section{Results:}
Though the four modules yield the four distinct results, we are attaching the results obtained from two modules, namely instance segmentation and the auto report generating module. The results of semantic segmentation and post-semantic segmentation (denoising part) are separately well demonstrated with superior scores for complex images by Das et al. \cite{das2024hip}. 

\begin{figure}[h!]
  \centering
  \begin{minipage}{0.24\textwidth}
    \centering
    \includegraphics[width=\textwidth]{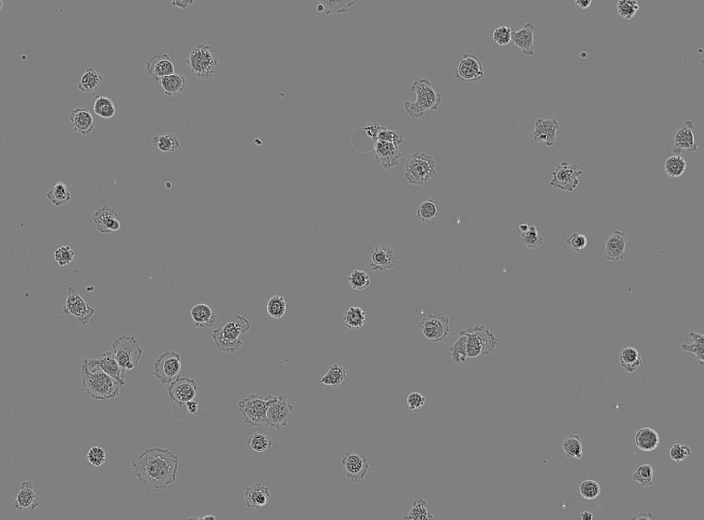}
    \vspace{0.01cm}
    \small\textbf{(a) Original input image.}
  \end{minipage}
  \hfill
  \begin{minipage}{0.24\textwidth}
    \centering
    \includegraphics[width=\textwidth]{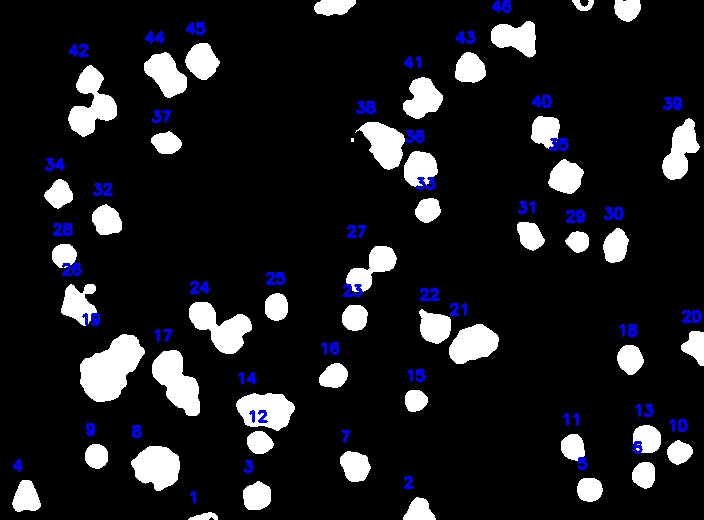}
    \vspace{0.01cm}
    \small\textbf{(b) Annotated cell boundaries.}
  \end{minipage}
 
  \vspace{0.1cm}
  
  \begin{minipage}{0.24\textwidth}
    \centering
    \includegraphics[width=\textwidth]{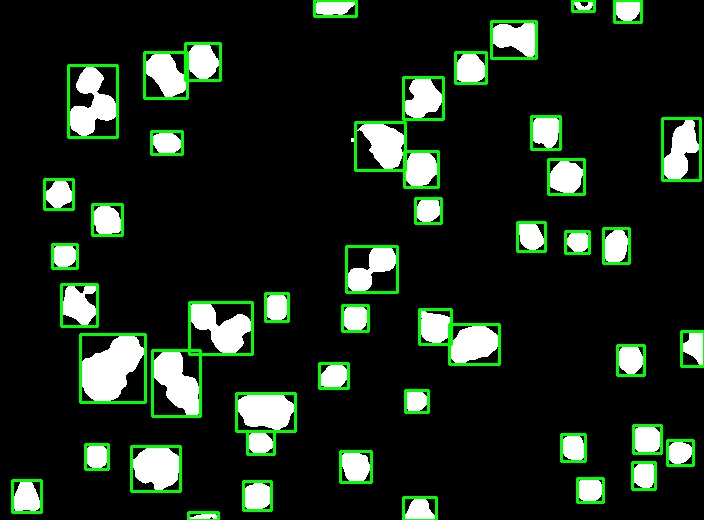}
    \vspace{0.01cm}
    \small\textbf{(c) Representation of bounding box overlay.}
  \end{minipage}
  \hfill
  \begin{minipage}{0.24\textwidth}
    \centering
    \includegraphics[width=\textwidth]{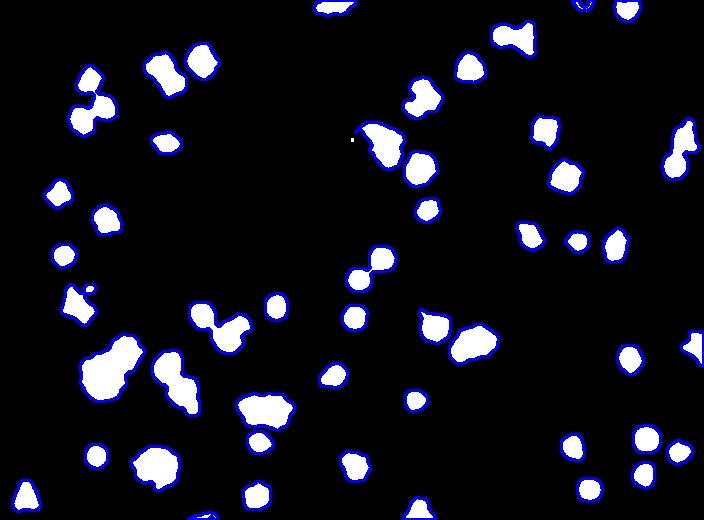}
    \vspace{0.01 cm}
    \small\textbf{(d) Detected contours showing individual cell instances.}
  \end{minipage}

  \begin{minipage}{0.24\textwidth}
    \centering
    \includegraphics[width=\textwidth]{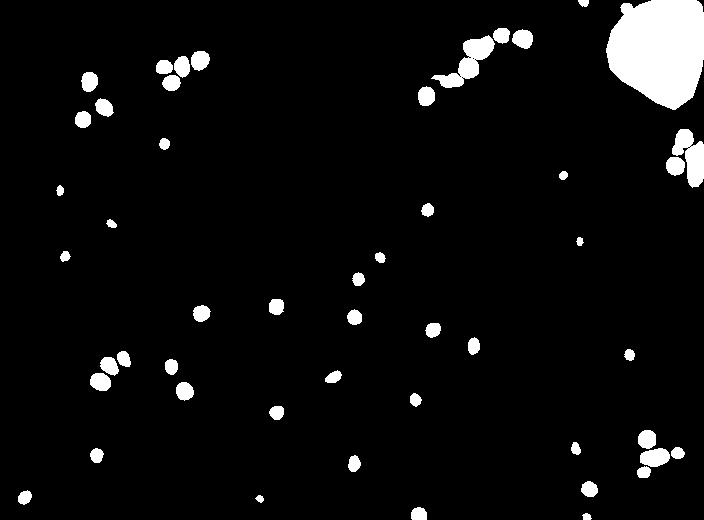}
    \vspace{0.01cm}
    \small\textbf{(e) Output by Stardist.}
  \end{minipage}
  \hfill
  \begin{minipage}{0.24\textwidth}
    \centering
    \includegraphics[width=\textwidth]{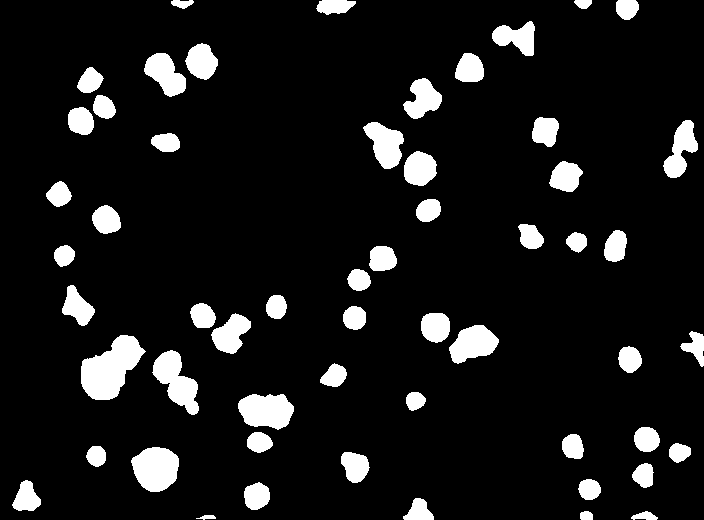}
    \vspace{0.01 cm}
    \small\textbf{(f) Output by Cellpose SAM}
  \end{minipage}
  
  \caption{{\color{black} (b), (c) and (d) are the results obtained from our segmentation and instance analysis modules applied to the LIVECell dataset (sample image: A172\_Phase\_A7\_1\_00d00h00m\_4.PNG). (e), (f) are segmentation by cellpose SAM and Stardist.}}
  \label{fig:image_processing_results}
\end{figure}

\begin{figure}[h!]
  \centering
  \begin{minipage}{0.24\textwidth}
    \centering
    \includegraphics[width=\textwidth]{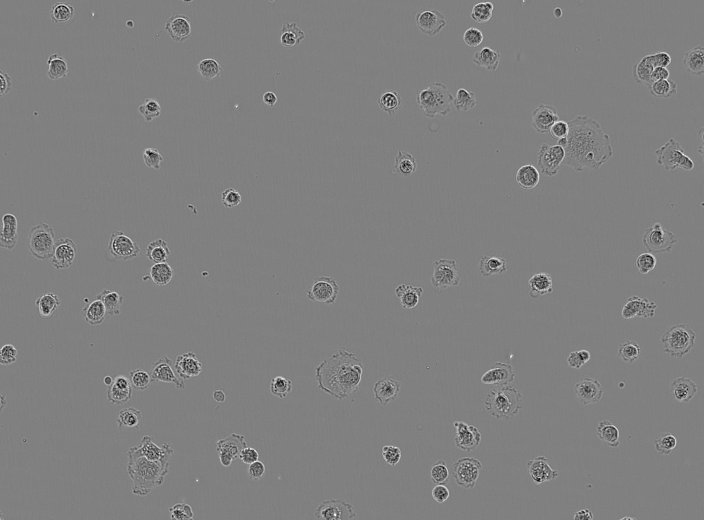}
    \vspace{0.01cm}
    \small\textbf{(a) Original input image.}
  \end{minipage}
  \hfill
  \begin{minipage}{0.24\textwidth}
    \centering
    \includegraphics[width=\textwidth]{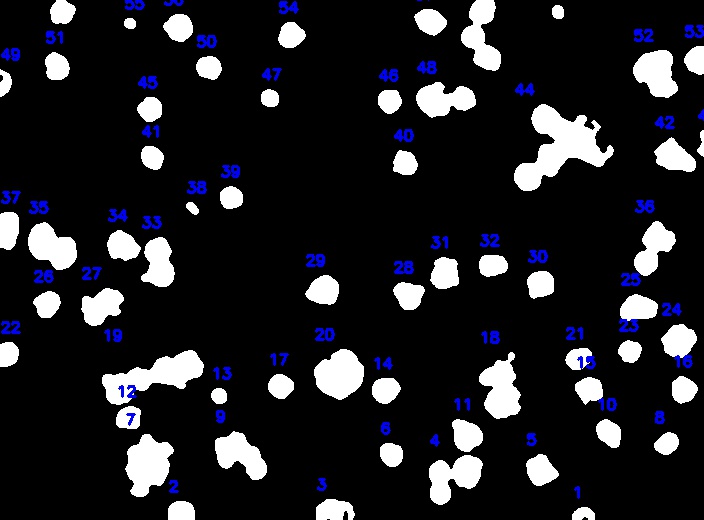}
    \vspace{0.01cm}
    \small\textbf{(b) Annotated cell boundaries.}
  \end{minipage}
 
  \vspace{0.1cm}
  
  \begin{minipage}{0.24\textwidth}
    \centering
    \includegraphics[width=\textwidth]{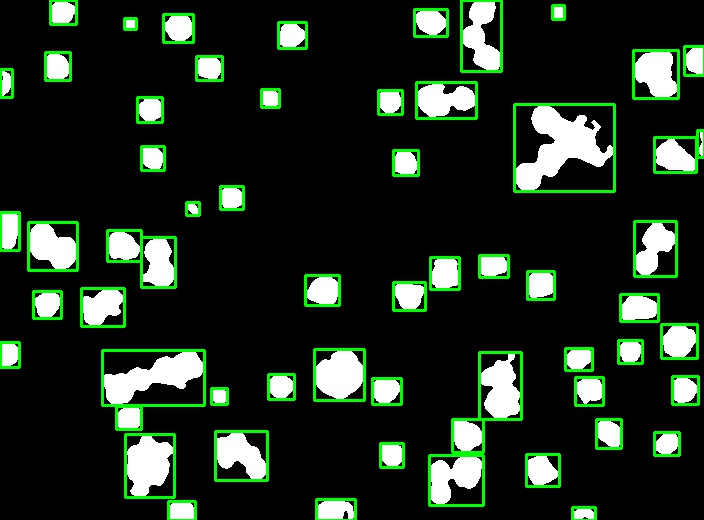}
    \vspace{0.01cm}
    \small\textbf{(c) Representation of bounding box overlay.}
  \end{minipage}
  \hfill
  \begin{minipage}{0.24\textwidth}
    \centering
    \includegraphics[width=\textwidth]{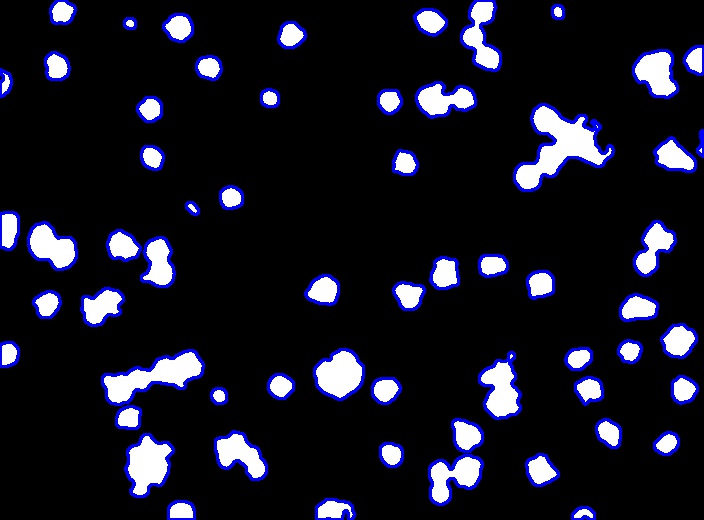}
    \vspace{0.01 cm}
    \small\textbf{(d) Detected contours showing individual cell instances.}
  \end{minipage}

  \vspace{0.1cm}
  
  \begin{minipage}{0.24\textwidth}
    \centering
    \includegraphics[width=\textwidth]{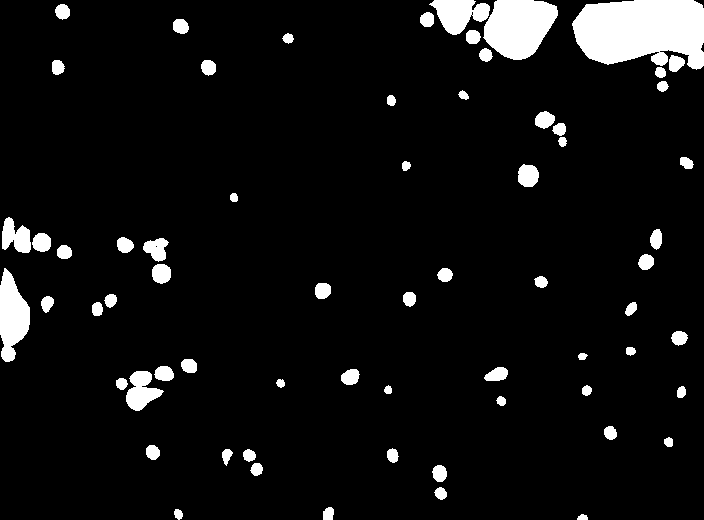}
    \vspace{0.01cm}
    \small\textbf{(e) Output by Stardist.}
  \end{minipage}
  \hfill
  \begin{minipage}{0.24\textwidth}
    \centering
    \includegraphics[width=\textwidth]{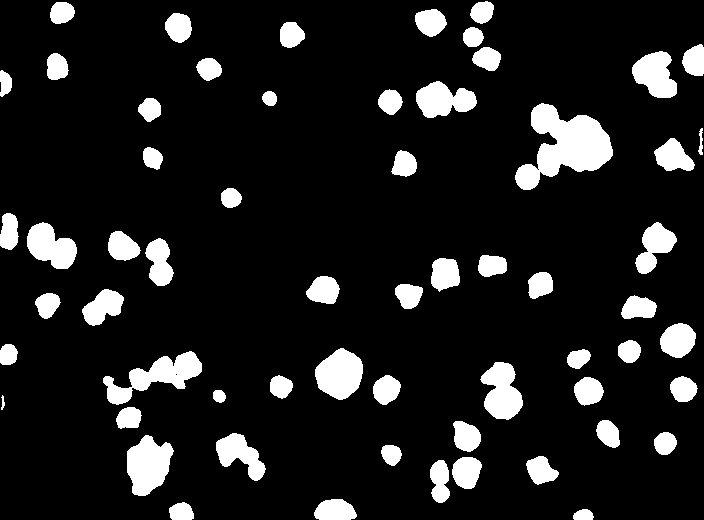}
    \vspace{0.01 cm}
    \small\textbf{(f) Output by Cellpose SAM}
  \end{minipage}
  
  \caption{{\color{black} (b), (c) and (d) are the results obtained from our segmentation and instance analysis modules applied to the LIVECell dataset (sample image: A172\_Phase\_A7\_1\_00d00h00m\_3.PNG). (e), (f) are segmentation by cellpose SAM and Stardist.}}
  \label{fig:image_processing_results_3}
\end{figure}

The first part of this section presents results from instance segmentation, preceded by semantic segmentation and post-segmentation denoising modules. {\color{black}Fig.~\ref{fig:image_processing_results} and \ref{fig:image_processing_results_3}} show outputs for two sample images from the public LIVECell dataset\footnote{\url{https://sartorius-research.github.io/LIVECell/}}
. {\color{black}Table~\ref{tab:gpu_comparison}} compares model performance across several metrics: CellPose-SAM processes images in ~4 s, CellPose3 and StarDist in ~3 s and ~4 s (semantic segmentation only, with I/O overhead), while our CPU-only end-to-end analysis system on an Intel NUC mini PC takes 20–30 s per image on average.

However, for different categories the scores differ. We mention the average scores of 3178 images {\color{black}in Table~\ref{tab:metrics2_transposed} with key metrics resulting from the algorithm validation. } Complete results are available at \footnote{ \url{https://drive.google.com/file/d/1UWfLJ1Zg263zppYRzza3OC9rI14q2qmO/view?usp=sharing}}.

\begin{table}[htbp]
\caption{Segmentation Performance Metrics on Dataset of 3,178 Images}
\centering
\begin{tabular}{|p{1.2cm}|p{2.2cm}|p{1.2cm}|p{2.2cm}|}
\hline
\textbf{Metric} & \textbf{Mean $\pm$ Std. Dev.} & \textbf{Metric} & \textbf{Mean $\pm$ Std. Dev.} \\
\hline
Dice & 0.8144 $\pm$ 0.0787 & Recall & 0.8512 $\pm$ 0.1246 \\
\hline
IoU & 0.6940 $\pm$ 0.1070 & F1 Score & 0.8144 $\pm$ 0.0787 \\
\hline
Accuracy & 0.8720 $\pm$ 0.0961 & SSIM & 0.5029 $\pm$ 0.2459 \\
\hline
Precision & 0.8034 $\pm$ 0.1126 & Hausdorff & 57.2874 $\pm$ 34.9202 \\
\hline
\end{tabular}
\label{tab:metrics2_transposed}
\end{table}

\begin{table*}[htbp]
\caption{Quantitative Comparison of Models on Two Sample Images}
\begin{center}
\begin{tabular}{|p{2.8cm}|p{1.4cm}|p{1.4cm}|p{1.4cm}|p{1.4cm}|p{1.4cm}|p{1.4cm}|p{1.8cm}|}
\hline
\textbf{Model} & \textbf{Dice} & \textbf{IoU} & \textbf{Accuracy} & \textbf{Precision} & \textbf{Recall} & \textbf{F1 Score} & \textbf{Hausdorff} \\
\hline
\multicolumn{8}{|c|}{\textit{Image: A172\_Phase\_A7\_1\_00d00h00m\_3.png}} \\
\hline
Das-Zun Model & 0.8942 & 0.8087 & 0.9727 & 0.8228 & 0.9792 & 0.8942 & 47.10 \\
\hline
CellPose-SAM & 0.9488 & 0.9026 & 0.9878 & 0.9364 & 0.9615 & 0.9488 & 69.18 \\
\hline
CellPose3 & 0.9493 & 0.9036 & 0.9878 & 0.9304 & 0.9691 & 0.9493 & 46.04 \\
\hline
StarDist & 0.3179 & 0.1890 & 0.8742 & 0.4395 & 0.2490 & 0.3179 & 91.92 \\
\hline
\multicolumn{8}{|c|}{\textit{Image: A172\_Phase\_A7\_1\_00d00h00m\_4.png}} \\
\hline
Das-Zun Model & 0.8851 & 0.7939 & 0.9763 & 0.8041 & 0.9843 & 0.8851 & 85.00 \\
\hline
CellPose-SAM & 0.9484 & 0.9018 & 0.9902 & 0.9281 & 0.9695 & 0.9484 & 45.69 \\
\hline
CellPose3 & 0.9429 & 0.8920 & 0.9891 & 0.9175 & 0.9698 & 0.9429 & 46.10 \\
\hline
StarDist & 0.3328 & 0.1996 & 0.9033 & 0.4606 & 0.2605 & 0.3328 & 129.42 \\
\hline
\end{tabular}
\label{tab:gpu_comparison}
\end{center}
\end{table*}

The second part of the result consists of several automated statistical quantifications and analyses of the image A172\_Phase\_A7\_1\_00d00h00m\_4.PNG.

\begin{figure}[htbp!]
  \centering
  \begin{minipage}{0.49\textwidth}
    \centering
    \includegraphics[width=\textwidth]{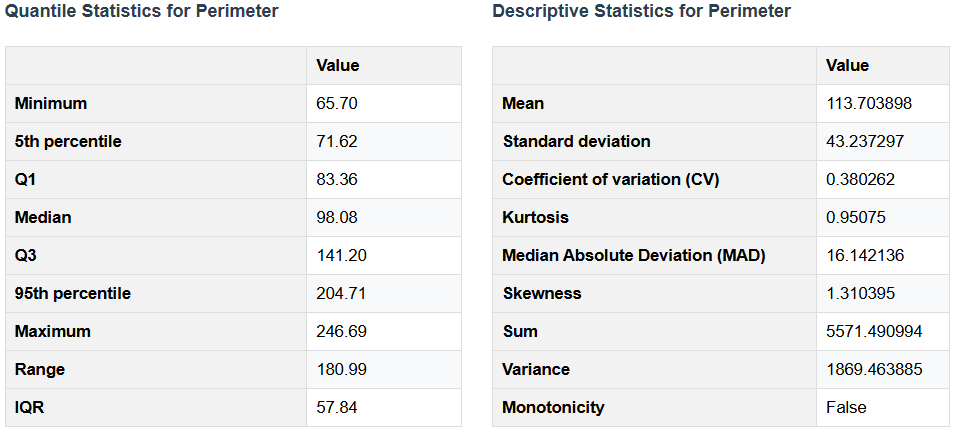}

  \end{minipage}
 
    \caption{Quartile statistics and descriptive statistics of Perimeter}
  \label{fig:Stat_pic}
\end{figure}

The system-generated cytometric dataset was processed through the automated reporting module to extract quantitative morphometric descriptors for 37 single-cell instances. The profiling yielded all numerical and categorical parameters, out of which a few have been chosen as the representative entities for this article: \textbf{Perimeter}, \textbf{Area}, \textbf{Aspect Ratio}, and \textbf{ID}, etc. exhibiting distinct distributional characteristics and inter-feature relationships. However, while more features (as described under Morphological Feature) could be selected based on user requirements, we limit our discussion to a few key ones along with some selective charts \& tables (Fig~\ref{fig:Stat_pic}, \ref{fig:hist_box}, \ref{fig:feature_correlation}) as the representative items in the following section:

\subsection{Descriptive Summary}

Summary statistics (Table~\ref{tab:summary_stats}) revealed that \textbf{Perimeter} and \textbf{Area} exhibited substantial variability, with mean values of $186.37 \pm 142.99$~µm and $1491.28 \pm 1310.80$~µm$^2$, respectively. The distributions were right-skewed (\textit{skew} = 3.02 for Perimeter; 2.49 for Area) and leptokurtic (\textit{kurtosis} = 9.81 and 6.90), suggesting the presence of a small subset of cells with disproportionately larger morphological dimensions. In contrast, \textbf{Aspect Ratio} maintained a narrow distribution centered at $1.00 \pm 0.30$, indicative of nearly circular cell geometries. No missing or duplicate entries were identified, confirming data completeness.

\subsection{Feature Correlation}

\begin{figure}[htbp]
\centering
\includegraphics[width=1\columnwidth]{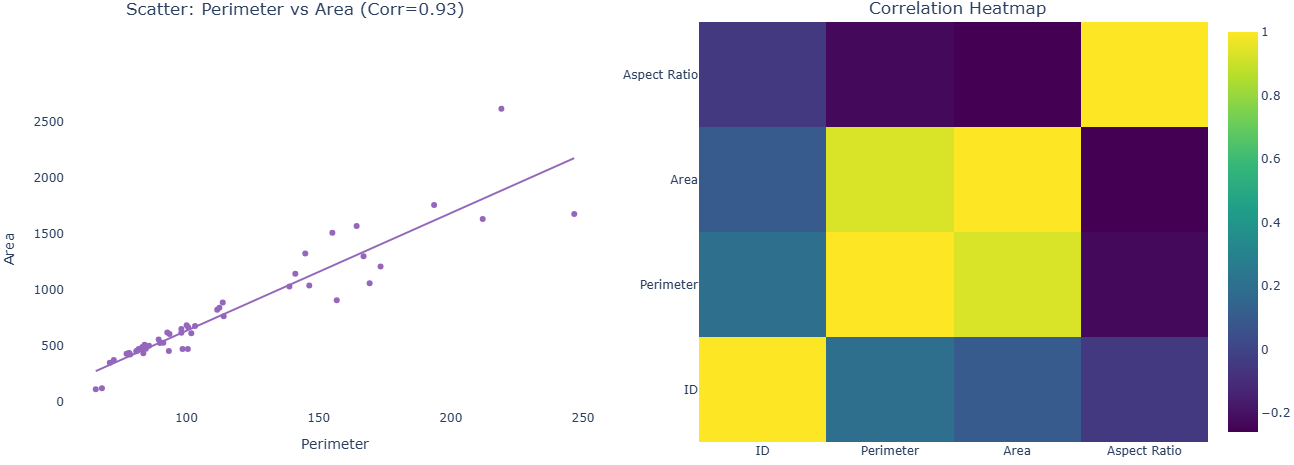}
\caption{(left)Feature correlation matrix showing a strong positive correlation between \textbf{Perimeter} and \textbf{Area} ($r = 0.97586$). (Right) Heatmap for the selected features}
\label{fig:feature_correlation}
\end{figure}

A strong positive correlation ($r = 0.97586$) was detected between \textbf{Perimeter} and \textbf{Area}, reflecting expected geometric proportionality (Fig~\ref{fig:feature_correlation}). This high correlation supports the consistency of the segmentation and measurement pipeline. The absence of other significant correlations indicates orthogonality between shape elongation (Aspect Ratio) and size-dependent variables, implying that elongation variations occurred independently of overall cell scale. Also, heatmap is generated with all selected features (Fig~\ref{fig:feature_correlation}).

\subsection{Distributional Analysis}

\begin{figure}[htbp]
\centering

\vspace{0cm}

\includegraphics[width=0.85\columnwidth]{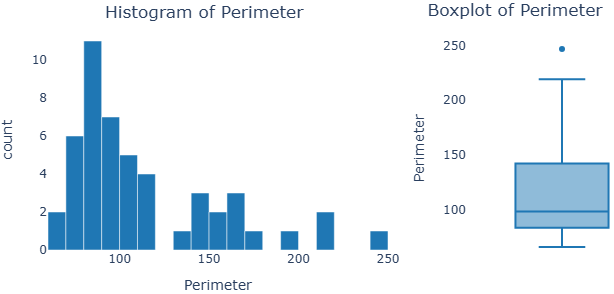}

\includegraphics[width=0.85\columnwidth]{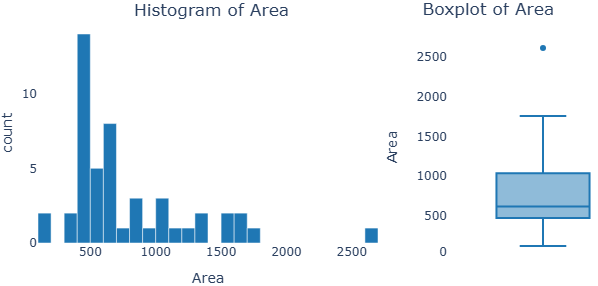}

\vspace{0cm}

\includegraphics[width=0.85\columnwidth]{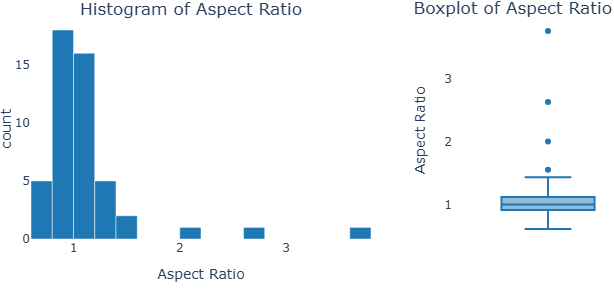}

\caption{Distribution and dispersion of morphometric parameter ( \textbf{Aspect ratio, }) across 37 cytometric samples. Histograms show the distribution of cell morphology features, while boxplots depict variation and outlier behavior.}
\label{fig:hist_box}
\end{figure}

Histograms and boxplots generated by the auto-report module (Figure~\ref{fig:hist_box}) demonstrated log-normal-like distributions for both Area and Perimeter, consistent with known biological variability in cytometric profiles. The interquartile ranges (IQRs) for Perimeter and Area were $87.74$~µm and $1027.00$~µm$^2$, respectively, suggesting moderate dispersion within the sample population. No outliers or negative values were detected, confirming robust preprocessing.

\subsection{Extreme Value and Frequency Analysis}

Extreme-value profiling identified a narrow subset of high-magnitude instances (Perimeter $>$ 420~µm; Area $>$ 4200~µm$^2$), each representing less than 3\% of the total population. Conversely, the smallest recorded objects (Area $<$ 300~µm$^2$) formed the lower distribution tail, potentially corresponding to debris or apoptotic fragments. Frequency-based inspection indicated that all measured instances were unique (100\% distinct), validating the single-cell sampling integrity.

\subsection{Interpretation}

Overall, the automated cytometric module accurately quantified morphometric variability within the sample, producing statistically consistent and biologically interpretable metrics. The high Perimeter--Area correlation validates internal feature coherence, while Aspect Ratio uniformity indicates minimal shape distortion during acquisition. These findings provide a reliable quantitative foundation for subsequent phenotype classification and feature selection analyses.

\begin{table}[htbp]
\caption{Summary Statistics of Representative Cytometric Features}
\begin{center}
\begin{tabular}{|p{2.2cm}|p{1.1cm}|p{1.1cm}|p{1.1cm}|p{1.1cm}|}
\hline
\textbf{Statistic} & \textbf{ID} & \textbf{Perimeter} & \textbf{Area} & \textbf{Aspect Ratio} \\
\hline
Count & 37 & 37 & 37 & 37 \\
\hline
Mean & 19.00 & 186.37 & 1491.28 & 1.00 \\
\hline
Standard Deviation & 10.82 & 142.99 & 1310.80 & 0.30 \\
\hline
Minimum & 1.00 & 80.18 & 225.50 & 0.42 \\
\hline
5th Percentile & 2.80 & 88.33 & 453.60 & 0.53 \\
\hline
25th Percentile (Q1) & 10.00 & 110.67 & 781.00 & 0.85 \\
\hline
Median (Q2) & 19.00 & 145.54 & 1100.50 & 1.00 \\
\hline
75th Percentile (Q3) & 28.00 & 198.41 & 1808.00 & 1.16 \\
\hline
95th Percentile & 35.20 & 421.37 & 4221.60 & 1.46 \\
\hline
Maximum & 37.00 & 769.98 & 6595.00 & 1.89 \\
\hline
\multicolumn{5}{p{6cm}}{$^{\mathrm{a}}$All measurements are derived from 37 single-cell observations. 
Perimeter is expressed in $\mu$m, Area in $\mu$m$^2$, and Aspect Ratio is dimensionless.} \\
\hline
\end{tabular}
\label{tab:summary_stats}
\end{center}
\end{table}

\section{Discussion}
{\color{black}
The software addresses a key gap in microscopic image analysis by providing a low-cost, label-free, unsupervised segmentation solution for live cells, with one-time calibration, auto analysis, and report generation, achieving competitive scores (often superior, as shown by Das et al. \cite{das2024hip}). Table~\ref{tab:gpu_comparison} shows all models can segment target objects with varying performance. Implementing Das \& Zun's algorithm, the software attains high Dice and F1 scores (0.885--0.894), comparable accuracy (0.972--0.976), and very high recall (0.979--0.984), indicating robust segmentation and sensitivity. Hausdorff distances confirm precise boundaries, while StarDist exhibits lower overlap and higher boundary errors. As demonstrated in \cite{das2024hip}, Das \& Zun's model performs well on complex images and often outperforms both CellPose and StarDist. Overall, these results highlight Das \& Zun's approach as a reliable, efficient, CPU-friendly, GUI-accessible alternative.}

\textbf{Broad Implications and Usability:} The GUI-driven application allows non-experts to perform advanced cytometry. Reproducible, batch-capable, and independent of cloud or GPU resources, it suits clinical diagnostics and cell therapy. Its efficacy on unstained cells aids long-term live-cell studies, and morphology-agnostic methods extend to bacterial colonies or tissue scaffolds.

\textbf{Limitations:} {\color{black}CPU performance lags GPU deep learning on large datasets. One-time calibration per imaging modality is required, and only 2D static images are supported; time-lapse and 3D analyses are not yet implemented.

\textbf{Future Directions:} Modular design allows optional lightweight deep learning modules for faster inference alongside the HIP model, 4D cell tracking, automated parameter tuning via heuristics or reinforcement learning, and community-driven support for diverse file formats and integration with platforms like Napari or OMERO. Additionally, this computer vision approach has the potential for an extentended research towards explainable AI (XAI) output}.

In summary, this software offers an accessible, versatile platform for label-free microscopy, with clear avenues for enhancement and adoption in biomedical research.

\section*{Conclusion}
{\color{black}The proposed software system democratizes advanced microscopy image analysis by offering a \textbf{low-cost, CPU-based, and training-free solution} for \textbf{label-free cytometric segmentation and auto-analysis}. It enables researchers—particularly in low-resource biomedical laboratories—to perform accurate and reproducible cell segmentation, feature extraction, and statistical reporting without the need for GPU hardware or programming skills.

The framework integrates four core components: (1) semantic and instance segmentation using fuzzy logic and spatial statistics; (2) post-segmentation refinement via morphological operations; (3) quantitative morphometric analysis for geometric and spatial characterization of cells; and (4) automated report generation using an interactive profiler. The unified pipeline supports both GUI and command-line execution, enhancing accessibility across user levels.

Compared with tools such as \textit{CellProfiler}, \textit{Cellpose}, and \textit{StarDist}, the proposed system provides a \textbf{fully local, unsupervised alternative} that removes dependence on training data, manual annotation, and high-performance computing. While deep learning approaches yield strong accuracy, their reliance on labeled datasets and GPUs restricts usability. In contrast, our CPU-based system achieves \textbf{comparable precision and reproducibility} on brightfield images of unstained cells.

By bridging the gap between algorithmic research and practical deployment, this software delivers a \textbf{reproducible and interpretable platform} for quantitative microscopic image analysis. Its adaptability to diverse morphologies and imaging modalities, along with automated reporting, put its value for \textbf{biomedical research}, including applications in \textit{cell transplantation}, \textit{regenerative medicine}, and \textit{tissue characterization}, where precise structural \& spatial quantification are essential}\footnote{This preprint is not peer-reviewed and is provided under a specific license. Unauthorized use is prohibited and is the sole responsibility of the user.}.

\section*{Disclosure of Interests}
The authors declare that they have no known competing financial interests or personal relationships
that could have appeared to influence the work reported in this paper.

\section*{Ethics Statement}
This study did not involve human participants or animals.

\section*{Author Contributions}
S.D. designed the study, analyzed the data, implemented the model, drafted the manuscript, and supervised the work. P.Z. provided supervision and critical review of the manuscript. All authors read and approved the final version

\end{document}